\documentclass[epj,nopacs]{svjour}
\usepackage{amsmath,amssymb,amsfonts,graphicx,cite,multirow}
\usepackage[utf8]{inputenc}
\graphicspath{{graphics/}}
\makeatletter
\ifx\input@path\@undefined
\def\input@path{{graphics/}}
\else
\g@addto@macro\input@path{{graphics/}}
\fi
\makeatother

\newcommand{\Hw}{\textsf{Herwig}}
\newcommand{\HWPP}{\textsf{Herwig++}}
\newcommand{\matchbox}{\textsf{Matchbox}}
\newcommand{\vbfnlo}{\textsf{VBFNLO}}
\usepackage{color}
\newcommand{\GeV}{\textrm{ GeV}}
\newcommand{\sq}[1]{\lq#1\rq}

\preprint{%
MAN/HEP/2016/08\\
HERWIG-2016-04\\
KA-TP-16-2016\\
MCnet-16-14\\
IPPP/16/49}

\title{Parton Shower Matching Systematics in Vector-Boson-Fusion WW Production}

\author{Michael Rauch\inst{1}\and Simon Pl\"atzer\inst{2,3}}

\institute{
Institute for Theoretical Physics, Karlsruhe Institute of Technology\and
Institute for Particle Physics Phenomenology, Durham University\and
School of Physics and Astronomy, University of Manchester}

\date{\today}

\abstract{We perform a detailed analysis of next-to-leading order plus
  parton-shower matching in vector-boson-fusion WW production including leptonic decays. The
  study is performed in the \textsf{Herwig}~7 framework interfaced to
  \textsf{VBFNLO}~3, using the angular-ordered and dipole-based parton-shower
  algorithms combined with the subtractive and multiplicative matching
  algorithms.}

\begin{document}

\maketitle


\section{Introduction}

Vector-boson fusion (VBF) and vector-boson scattering (VBS) are an important
class of processes for the Large Hadron Collider (LHC) at CERN. These
processes are characterised by the electroweak production of single bosons and
di-bosons in association with two jets, respectively, where a quark or
anti-quark scatters with another quark or anti-quark through the space-like
exchange of an electroweak boson.  VBF and VBS production are particularly
important for the recently started run-II phase of the LHC, as their cross sections
significantly increase due to the higher centre-of-mass energy of 13~TeV.
Its study was first suggested for the VBF production of Higgs
bosons~\cite{Cahn:1983ip,Dawson:1984gx,Duncan:1985vj,Cahn:1986zv,Kleiss:1987cj,Barger:1988mr,Butterworth:2002tt,Dittmaier:2011ti,Dittmaier:2012vm,Heinemeyer:2013tqa}.
In the following, to simplify notation we will collectively refer to both
types of processes as VBF.

The characteristic feature of the VBF class of processes are two energetic
jets in the forward regions of the detector, the so-called tagging
jets~\cite{Zeppenfeld:1999yd}. In the central region, only a low jet activity
is observed. The leptonic decay products of the vector bosons are typically
found between the two tagging jets.  These properties allow us to distinguish
VBF from two types of background processes with the same final state. At the
same order in the coupling constants, di-boson or tri-boson production, where
one of the bosons decays hadronically,
contributes~\cite{Lazopoulos:2007ix,Hankele:2007sb,Campanario:2008yg,Binoth:2008kt,Bozzi:2009ig,Baur:2010zf,Bozzi:2010sj,Bozzi:2011wwa,Bozzi:2011en,Feigl:2013naa}.
For these processes, the invariant mass of the two jets is close to the
mass of the decaying boson, and larger values are strongly suppressed. The
other class of irreducible backgrounds is QCD-induced production in
association with two
jets~\cite{Melia:2010bm,Melia:2011dw,Melia:2011gk,Jager:2011ms,Greiner:2012im,Campanario:2013qba,Gehrmann:2013bga,Campanario:2013gea,Badger:2013ava,Campanario:2014dpa,Bern:2014vza,Alwall:2014hca,Campanario:2014ioa,Campanario:2014wga,Campanario:2015vqa}.
There, the two jets are preferably emitted in the central region.  As two
powers of the electromagnetic coupling constant get replaced by their strong
counterpart, and for most boson combinations gluon-induced production
channels are also possible, this production mode will dominate for inclusive cross
sections.  Applying tight VBF cuts allows us to reduce these background
processes and suppress any interference effects~\cite{Campanario:2013gea}.
These cuts typically require a large invariant mass of the two tagging jets of
the order of several hundreds GeV, and a large rapidity separation between
them. They also reduce any interference effects between $t$- and $u$-channel
exchange diagrams to a completely negligible level and justify the often used
so-called VBF approximation or structure-function approach, where these
contributions are not taken into account.  A veto on additional
jets~\cite{Rainwater:1996ud} can further enhance the signal-to-background
ratio.

The appearance of triple and, in the case of di-boson production, quartic
gauge-boson vertices makes VBF processes an ideal tool for studying these. In
the high-energy region, a strong cancellation between diagrams with quartic
vertices, triple vertices, and Higgs boson exchange takes place. Any
modifications of the couplings from their Standard Model (SM) values could spoil
this cancellation and lead to a rise of the squared matrix element
proportional up to the eighth power of the di-boson invariant mass. VBF
is therefore a sensitive probe of these anomalous contributions to the SM gauge
couplings. Also, the existence of additional heavy Higgs bosons from
additional singlets or doublets can be probed in the di-boson invariant-mass
distribution~\cite{Heinemeyer:2013tqa,LHCHXSWGYR4}.

To investigate such effects of physics beyond the Standard Model (BSM), a
precise knowledge of the underlying SM prediction is
necessary. Next-to-leading order (NLO) QCD corrections to all VBF processes
have been
computed~\cite{Figy:2003nv,Oleari:2003tc,Jager:2006zc,Jager:2006cp,Bozzi:2007ur,Jager:2009xx,Denner:2012dz,Campanario:2013eta}.
Their effect is typically rather modest, of the order of 10\% or below.
Choosing the momentum transfer through the space-like bosons as a scale choice
has been proven to be a very good choice.  A dedicated implementation of all
VBF processes at NLO QCD accuracy, including leptonic decays of the vector
bosons and the option to switch on anomalous coupling effects or some BSM
models such as a two-Higgs model, is available in the \vbfnlo{}
program~\cite{Arnold:2008rz,Arnold:2011wj,Arnold:2012xn,Baglio:2014uba}.

The combination of NLO QCD results with parton showers has been studied thus far
for some of the VBF
processes~\cite{Nason:2009ai,Jager:2011ms,Jager:2012xk,Schissler:2013nga,Jager:2013mu,Jager:2013iza,Jager:2014vna}
using the POWHEG-BOX
framework~\cite{Nason:2004rx,Frixione:2007vw,Alioli:2010xd}.  Additional
corrections have so far been calculated only for VBF-$H$ production. These are
the NLO electroweak
corrections~\cite{Ciccolini:2007jr,Ciccolini:2007ec,Figy:2010ct}, which turn
out to be of a similar size to the NLO QCD ones, but for the measured Higgs
mass are of opposite sign. Also known are the NLO QCD corrections for VBF-$H$
production in association with three jets
\cite{Figy:2007kv,Campanario:2013fsa}. A third type are the
next-to-next-to-leading order QCD corrections, using the structure-function
approach. While corrections to the inclusive cross section are well below the
percent level~\cite{Bolzoni:2010xr,Bolzoni:2011cu}, there are much larger effects 
for differential distributions~\cite{Cacciari:2015jma}. These are, however,
mostly due to the additional effects from double real-radiation processes, and
are reasonably described by adding parton-shower effects on top of NLO QCD results.

For a detailed understanding of VBF processes the matching of NLO
QCD predictions with parton showers is therefore necessary. This includes not only the
central predictions, but also trying to quantify the associated theory
errors. Tools to assess them are for example the variation of various scales
appearing in the predictions. However, one can also compare different matching and
parton-shower algorithms. Combining the fast and accurate predictions of
\vbfnlo{} with the flexible options of \Hw~7 hence offers us unique
possibilities to study these effects. As it is important to have control of these
uncertainties in the perturbative part of the simulation, we
will not consider any effects due to hadronisation or multiple parton
interactions. These are left for a future publication.

\section{Outline of the Simulation}

\subsection{NLO and NLO+PS Matching with Herwig 7}

The newly released \Hw~7 Monte Carlo event generator \cite{Bellm:2015jjp,hwweb} 
builds on its successful predecessor \HWPP~\cite{Bahr:2008pv}. It features
significantly improved physics capabilities, particularly for 
NLO QCD corrections and their combination with the two available
parton shower modules based on Refs.~\cite{Gieseke:2003rz}
and~\cite{Platzer:2009jq}.

Based on extensions of the previously developed \matchbox\ module
\cite{Platzer:2011bc}, NLO event simulation can be carried out with the help
of external amplitude providers, which are used by \Hw\ to evaluate tree-level
and one-loop matrix elements. These are then automatically combined with the
Catani-Seymour dipole subtraction \cite{Catani:1996vz,Catani:2002hc}, and
general-purpose as well as specialised phase space generation algorithms to
assemble a full NLO calculation. This NLO calculation can be further extended
by the automatically determined matching subtractions to combine it with a
downstream parton shower algorithm.  While a number of hard process
calculations are supported by dedicated \matchbox\ plugins, communication with
external general-purpose amplitude providers, \vbfnlo~3
\cite{Arnold:2008rz,Baglio:2014uba,vbfnloweb} in the context of this
study, takes place via extensions of the BLHA 2 standard
\cite{Alioli:2013nda}.

NLO predictions obtained from the \Hw $+$\vbfnlo\ setup have extensively
been validated against standalone calculations obtained from \vbfnlo, using
both a range of integrator and phase space generation algorithms either
supplied by the standard \matchbox\ modules or employing the versatile interface
structure to use the according \vbfnlo\ routines.

\subsection{VBFNLO 3}

\begin{figure}
\begin{center}
\includegraphics[width=0.45\columnwidth]{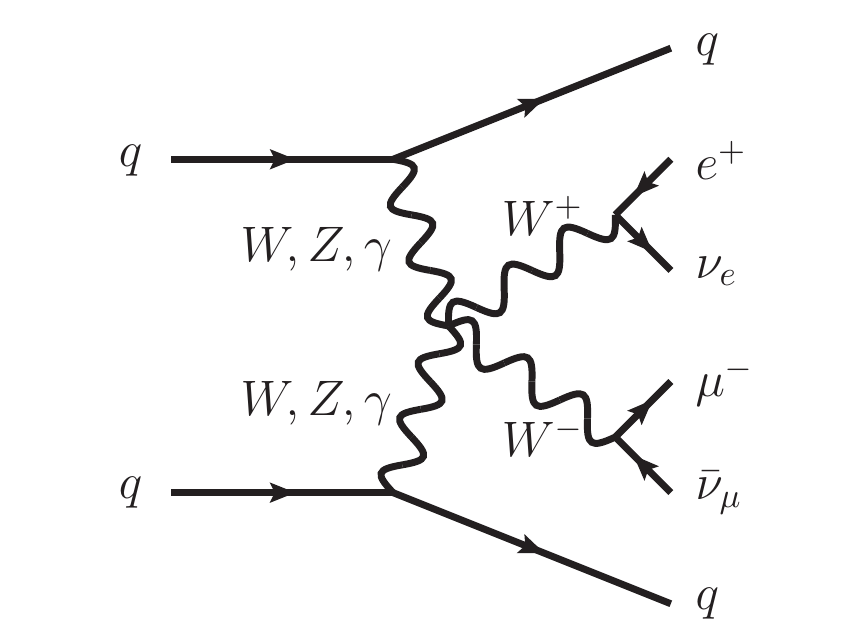}
\includegraphics[width=0.45\columnwidth]{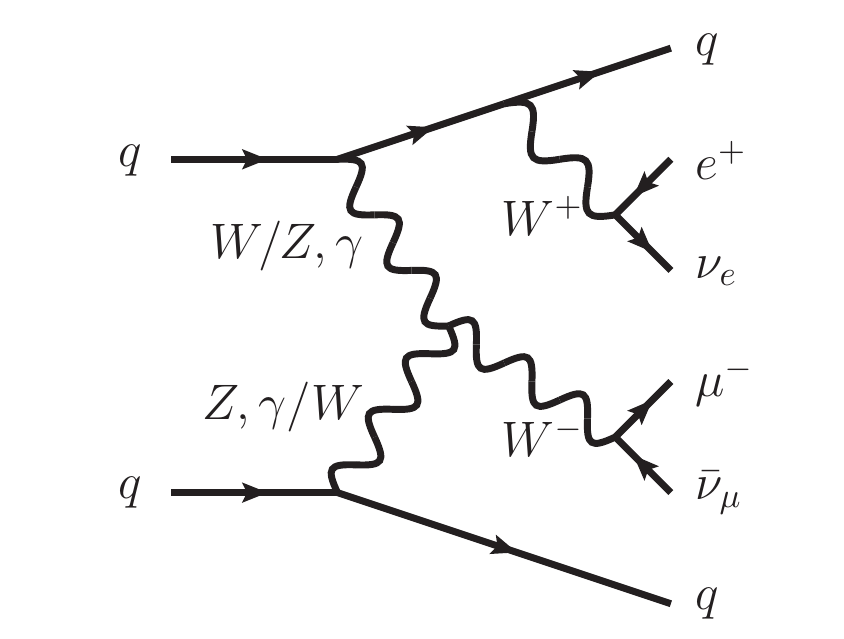} \\[1ex]
\includegraphics[width=0.45\columnwidth]{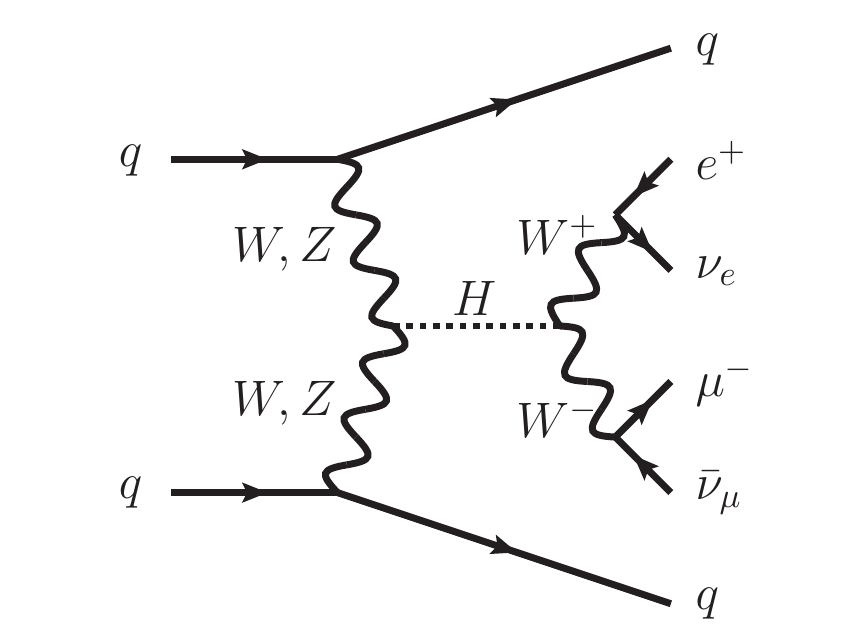}
\includegraphics[width=0.45\columnwidth]{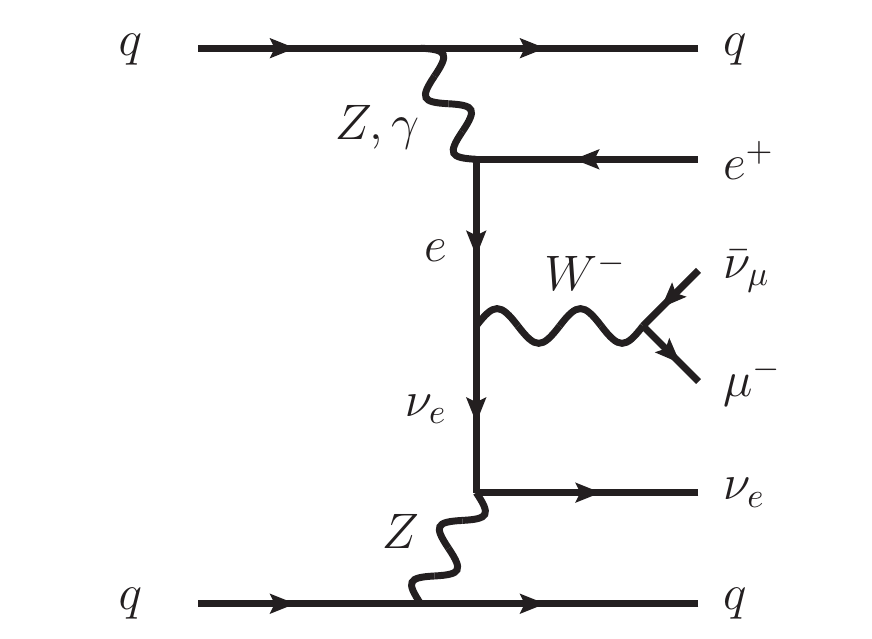} 
\end{center}
\caption{Example Feynman diagrams for VBF-$W^+W^-jj$ production at LO including
leptonic decays and non-resonant contributions.}
\label{fig:feynman}
\end{figure}

\vbfnlo~\cite{Arnold:2008rz,Baglio:2014uba,vbfnloweb} is a flexible parton-level
Monte Carlo generator for processes with electroweak bosons. It provides a fully
differential simulation of VBF processes, amongst others, at NLO QCD
accuracy.

The communication with \Hw\ is done via an interface based on the
BLHA 2 standard \cite{Alioli:2013nda}. The \vbfnlo\ implementation
contains some extensions on top of the standard to access additional
features such as the internal phase-space generator, which has been used for
the results presented in this article. None of them is mandatory,
however, and a standards-compliant Monte Carlo generator is sufficient
to obtain amplitudes from \vbfnlo.

In this article, we are considering as an example the electroweak
production process 
\begin{equation}
pp \rightarrow W^+ W^- jj \rightarrow e^+ \nu_e \mu^- \bar{\nu}_\mu jj \,.
\end{equation}
We include the leptonic decays of the $W$ bosons including full
off-shell effects. Contributions from non-resonant diagrams as well as
those with Higgs bosons are included. The latter are not only
important in phase-space regions where the Higgs boson becomes on-shell, but
also at large invariant masses of the four leptons, where a significant
cancellation between Higgs boson and continuum diagrams takes
place~\cite{Heinemeyer:2013tqa,LHCHXSWGYR4}. For the partons we restrict ourselves to the VBF
approximation, where interference effects between same-flavour quarks in
the final-state are neglected. These terms are both phase-space and
colour-suppressed. When imposing VBF-specific cuts, their contribution
to the cross section becomes negligible.
Some example Feynman diagrams of the LO process are depicted in
Figure~\ref{fig:feynman}. They show the rich structure of this process,
which includes contributions from quartic gauge-boson vertices
(\textit{top left diagram}), triple gauge-boson vertices (\textit{top
right}), and Higgs-mediated exchange (\textit{bottom left}). An example
of a non-resonant contribution is shown in the bottom right diagram of
Figure~\ref{fig:feynman}.
The NLO QCD corrections to this process have been first presented
in Ref.~\cite{Jager:2006zc}.

\section{Matching Algorithms and Uncertainties}

\matchbox\ currently supports direct, subtractive matching ({\it i.e.}
MC@NLO-type \cite{Frixione:2002ik}) to both the angular ordered and dipole
showers, as well as multiplicative ({\it i.e.} Powheg-type
\cite{Nason:2004rx}) matching. Conceptually, as well as technically, these
algorithms are calculating matched cross sections as\footnote{Further details
  to the matching and the other algorithms provided by \Hw\ will be subject to
  an extensive discussion in an upcoming review.}
\begin{eqnarray}
  \sigma_{\text{NLO}}^{\text{matched}} &=& \int_n \left( {\rm
    d}\sigma_{\text{LO}} + {\rm d}\sigma_{\text{virt}}\right) \\\nonumber &+&
  \int_n \int_1 \left( {\rm d}\sigma_{\text{PS}}- {\rm
    d}\sigma_{\text{sub}}\right)\\\nonumber &+& \int_{n+1}\left( {\rm
    d}\sigma_{\text{R}}- {\rm d}\sigma_{\text{PS}}\right) \ ,
\end{eqnarray}
where in this schematic notation \sq{LO} denotes the leading order cross section,
\sq{virt} the contribution by one-loop diagrams, integrated subtraction terms and
collinear counter-terms, \sq{sub} denotes un-integrated subtraction terms, \sq{R}
the real emission and \sq{PS} is the parton-shower approximation to the
real-emission cross section. The second integral in the middle line is
performed over the one-particle phase space of the extra emission.
In this notation, the parton shower approximation
can also be given by a matrix element correction
\cite{Seymour:1994df,Miu:1998ju}, giving rise to multiplicative, or Powheg-type,
matching. \matchbox\ samples the matrix-element corrections using adaptive
methods \cite{Platzer:2011dr} and is able, for the case of the angular-ordered
shower, to add truncated showers on top of it to fully account for
large-angle, soft emissions.

Uncertainties are explored by varying the relevant scales in the hard process
and showers as outlined in Ref.~\cite{Bellm:2016rhh}. For both of the showers, as
well as for the matrix-element correction entering the multiplicative matching
we choose to use the \texttt{resummation} profile scale~\cite{Bellm:2016rhh} to guarantee a smooth
transition between the hard matching and resummation regions, while
maintaining the resummation properties of the parton shower.

\section{Phenomenological Results}

We perform parton-level studies, treating all partons as massless. As we
are not interested in effects from top-quark production, we apply a veto
on any bottom quarks appearing in the final state.
Multiple parton interactions (MPI) are not included, and showering
is performed using both \Hw\ shower modules at their default settings. We also
employ default settings for the hard process calculations, including the
MMHT2014 PDF set \cite{Harland-Lang:2014zoa} with five active flavours.

We apply typical VBF selection cuts,
\begin{align}
p_{T,j} &> 30 \GeV \,, & |y_j| &< 4.5 \,, \nonumber\\
p_{T,\ell} &> 20 \GeV \,, & |y_{\ell}| &< 2.5 \,, \nonumber\\
m_{e^+,\mu^-} &> 15 \GeV \,, \nonumber\\
m_{j1,j2} &> 600 \GeV \,, & |y_{j1}-y_{j2}| &> 3.6 \,,
\label{eq:vbfww_cuts}
\end{align}
and consider $pp$ collisions at a centre-of-mass energy of $13\ {\rm
  TeV}$. Jets are clustered from partons using the anti-$k_T$
algorithm~\cite{Cacciari:2008gp} with a cone radius of $R=0.4$.
The choice of cuts is adopted from the cut-based VBF category of the
$H\rightarrow WW$ study of ATLAS~\cite{ATLAS:2014aga}. The corresponding CMS
analysis~\cite{Chatrchyan:2013iaa} uses very similar
values.
We take the transverse momentum of the
leading jet, $\mu_0 = p_{T,j1}$, as the central scale choice. 

\begin{figure}
\begin{center}
\includegraphics[scale=0.55]{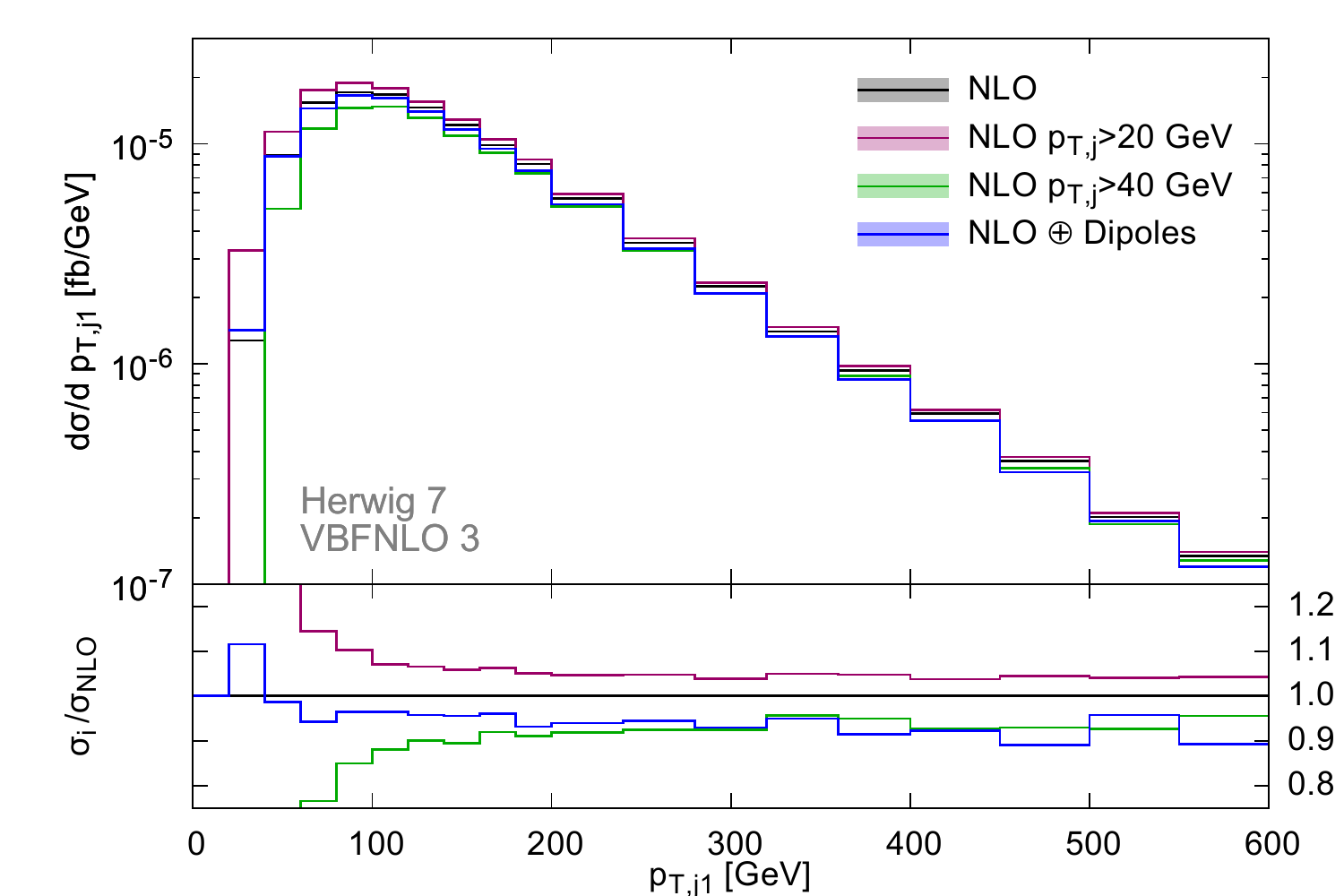}
\end{center}
\caption{\label{figures:cutmigration}The transverse-momentum spectrum of the
  leading jet for a number of different cuts on the fixed order calculation
  comparing the central NLO and showered result using a transverse momentum
  threshold of $30\ {\rm GeV}$. Cut migration effects for the central
  predictions at high transverse momenta are at the level of $10 \%$, while
  they can have a significant impact in lower-$p_\perp$ observables.}
\end{figure}

Processes with jets at the level of the hard process require selection cuts
on the jets; additional parton shower emissions off these jets will migrate
contributions across the cut boundary such that jet cross sections after
applying parton showering will typically be lower than the input
cross section at the level of the hard process. We quantify this effect in
Fig.~\ref{figures:cutmigration} by sliding the jet cut at the level of the
fixed order NLO cross section from $p_\perp \ge 20,30$ through $40\ {\rm
  GeV}$ and comparing to the showered result using a jet $p_\perp$ threshold of
$30\ {\rm GeV}$. As representative observable we take the
$p_\perp$ spectrum of the leading jet in this case, though similar findings
apply to the other observables and inclusive cross sections, as well. Choosing
the analysis cut to be equal to the generation cut is well contained within
the variation of the cut applied at the hard process. We therefore conclude
that no further tuning of acceptance criteria to minimise cut migration is
required in this study.
To err on the side of caution, we nevertheless apply generation-level
cuts which are looser than the ones given in eq.~\ref{eq:vbfww_cuts}.
An event is selected for further processing if at least two jets with
transverse momenta larger than $20 \GeV$ within a rapidity range of
$|y|<5$ are present, and the two leading jets have an invariant mass of
at least $400 \GeV$ with a rapidity separation larger than $3$. Also the
lepton cuts are relaxed to a minimum transverse momentum of $15 \GeV$
and an absolute value of the rapidity smaller than 3.

Contrary to the study presented in Ref.~\cite{Bellm:2016rhh}, here we have 
considered the parton showers at their (tuned) default settings rather than
the baseline settings; we expect the effects caused by these differences to be
small. The only noticeable difference in variations is a larger down-variation
of the angular-ordered shower when lowering the renormalization scale
appearing as argument of the strong coupling; this effect is only visible at
the level of the hard tagging jets and we therefore conclude that it is
originating from an increased cut migration due to enhanced radiation present
in this variation.

\begin{figure}
\begin{center}
\includegraphics[scale=0.55]{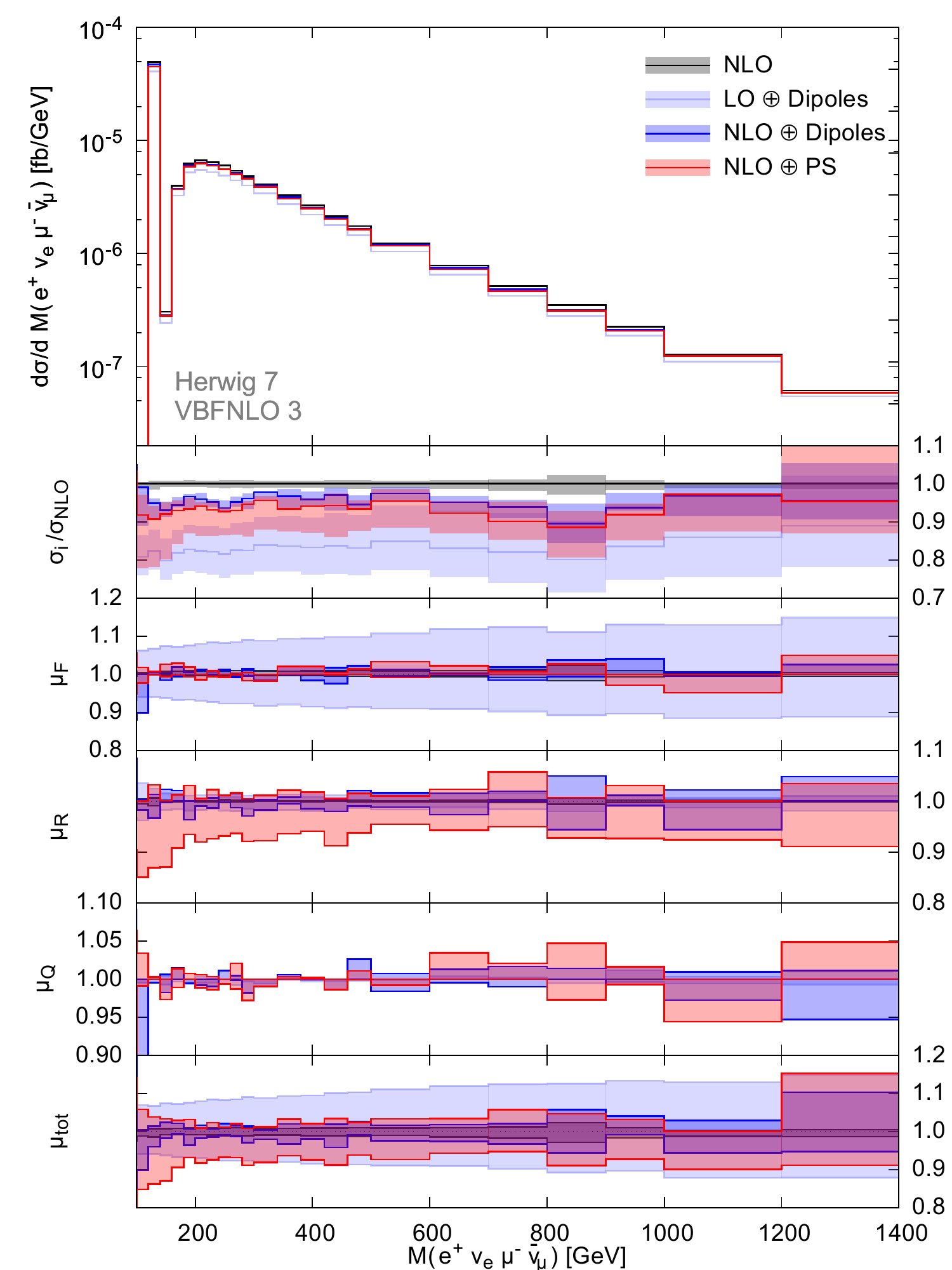}
\end{center}
\caption{\label{figures:WWMass}The invariant 4-lepton mass comparing parton-level
  NLO results (black) with scale variations, leading order plus dipole shower
  predictions (light blue), and NLO matched results for the dipole and angular-ordered
  shower (dark blue and red, respectively). The top ratio plot shows the
  central predictions and overall variation envelopes with respect to the
  parton-level fixed-order result; the subsequent ratio plots show the
  variations of the individual scales with respect to their central
  predictions, focusing on factorization ($\mu_F$), renormalization ($\mu_R$)
  and hard veto scale ($\mu_Q$) variations, as well as the overall envelope
  ($\mu_{\text{tot}}$).}
\end{figure}

Turning to uncertainties we first consider the distribution of the
four-lepton invariant mass depicted in Fig.~\ref{figures:WWMass}. 
The larger upper panel shows the differential
distributions using the central scale choice, exhibiting the Higgs boson
peak at $125\ {\rm GeV}$ and the continuum production region above $2
M_W$. Curves shown are the parton-level NLO results (black), leading
order plus dipole shower (light blue) and NLO matched results for the
dipole and angular-ordered shower (dark blue and red, respectively). The
uppermost of the smaller panels shows the ratio of the cross section
with respect to the parton-level fixed-order result, while the bands
depict the overall scale variation envelopes. The four lower panels show
the changes of the differential cross section when varying, from top to
bottom, the factorization ($\mu_F$), renormalization ($\mu_R$) and hard
veto scale ($\mu_Q$), and all of them ($\mu_{\text{tot}}$). Variations
are performed in the range $\mu_i/\mu_0 \in [\frac12;2]$. For the total
uncertainty envelope, we allow the individual scales to vary
independently, but require that ratios of scales also fulfil the
condition $\mu_i/\mu_j \in [\frac12;2]$. 

\begin{figure}
\begin{center}
\includegraphics[scale=0.55]{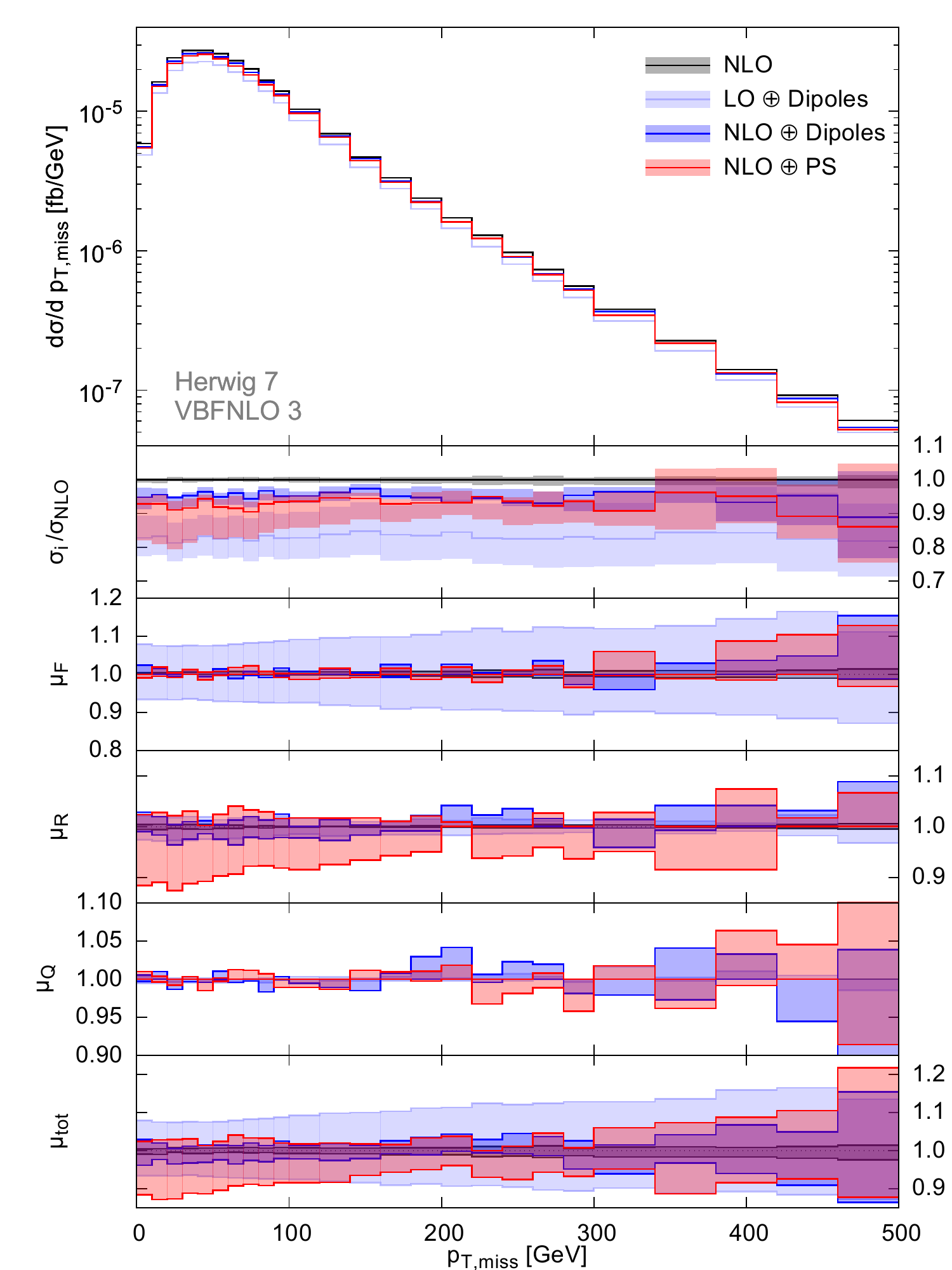}
\end{center}
\caption{\label{figures:PtMissPt}The cross section predictions differential in
  the missing transverse momentum. See Fig.~\ref{figures:WWMass} and the text
  for more discussion.}
\end{figure}

\begin{figure}
\begin{center}
\includegraphics[scale=0.55]{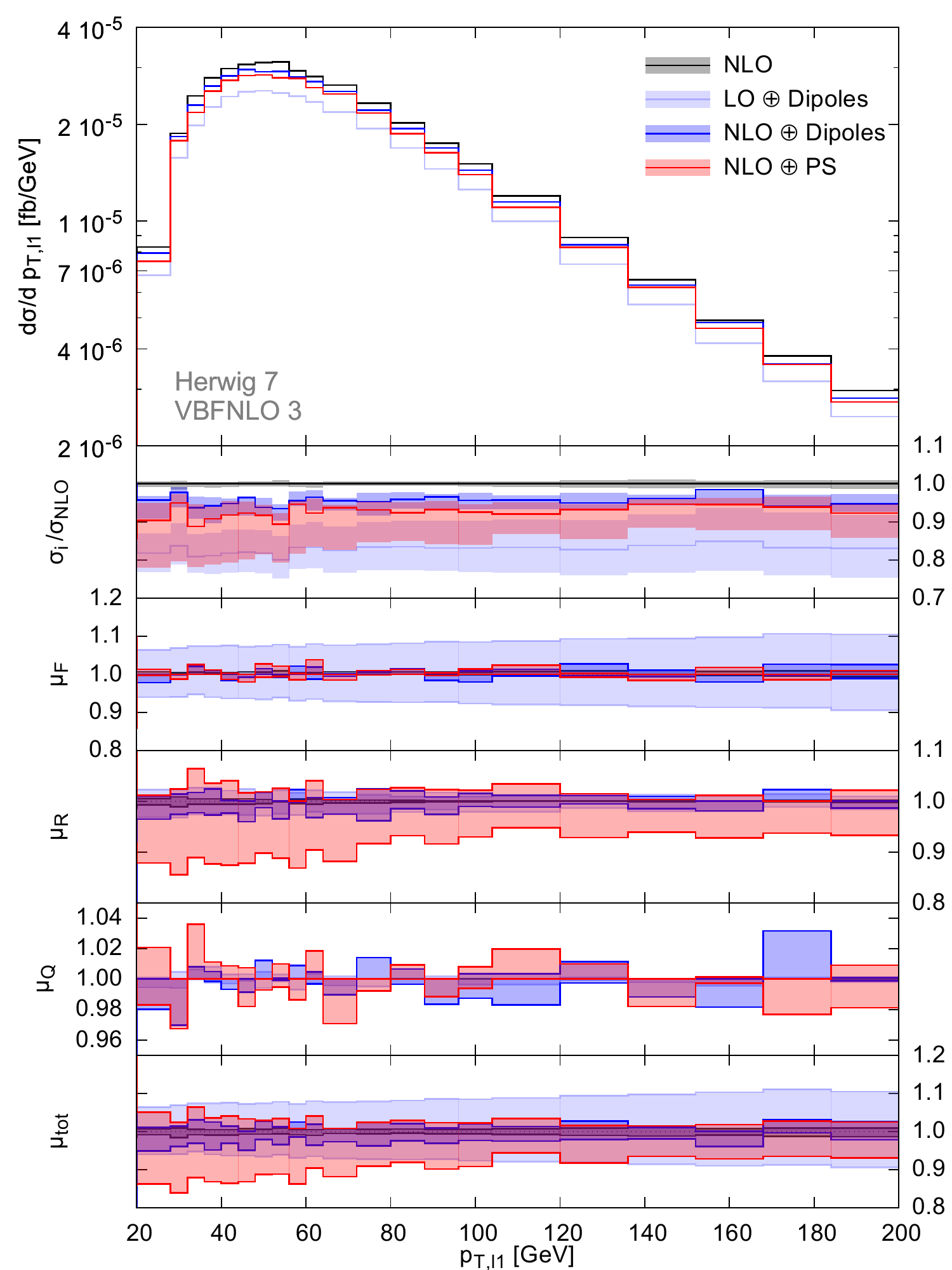}
\end{center}
\caption{\label{figures:ChargedLepton1Pt}Same as Fig.~\ref{figures:WWMass},
  showing the $p_\perp$ spectrum of the leading charged lepton. The spectrum
  of the subleading lepton shows a similar behaviour.}
\end{figure}

We find that parton showering only mildly affects the shape of the
four-lepton invariant mass distribution, while the overall normalisation
is subject to
configurations showered 'out' of the VBF acceptance criteria. The shower
uncertainties are clearly reduced in changing from LO+PS to NLO+PS simulation,
with both showers yielding comparable results both in their central prediction
as well as variations. Similar conclusions apply to other observables probing
mainly the electroweak part of the final state, such as the missing transverse
momentum distribution Fig.~\ref{figures:PtMissPt} and the $p_\perp$ spectrum
of the leading charged lepton Fig.~\ref{figures:ChargedLepton1Pt}.

\begin{figure}
\begin{center}
\includegraphics[scale=0.55]{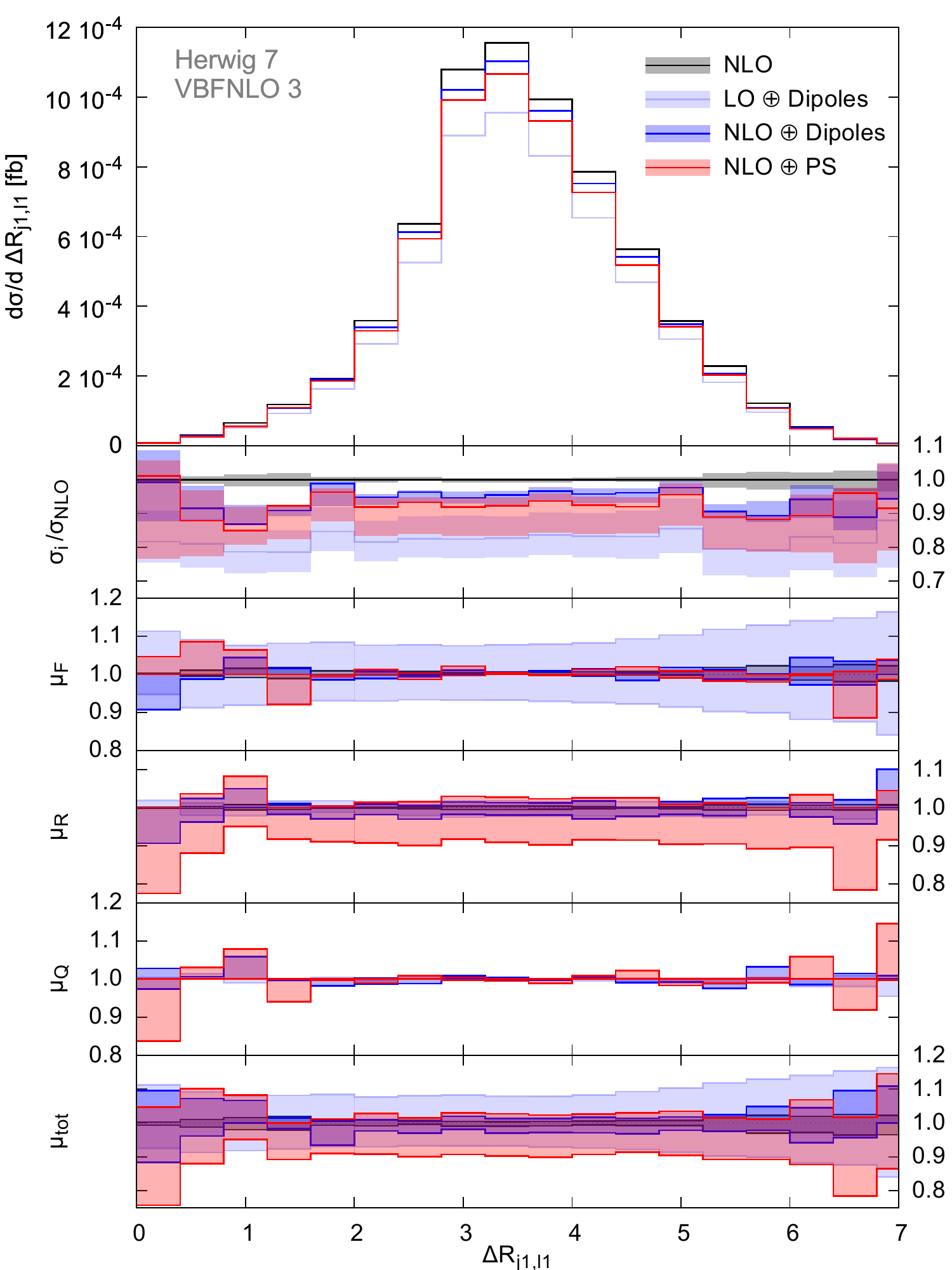}
\end{center}
\caption{\label{figures:LeptonSeparation}Same as Fig.~\ref{figures:WWMass},
  showing the separation of the leading charged lepton and the leading
  jet. Separations with respect to the subleading, second and third, jets show
  similar features.}
\end{figure}

Further observables required to reconstruct the VBF signature are significantly
more affected by parton shower effects, exemplified here in the case
of the separation between the leading lepton and the leading jet shown in
Fig.~\ref{figures:LeptonSeparation}, with the separation from the third jet
being most sensitive to shower effects; within uncertainties, the showers do,
however, yield comparable results.

\begin{figure}
\begin{center}
\includegraphics[scale=0.55]{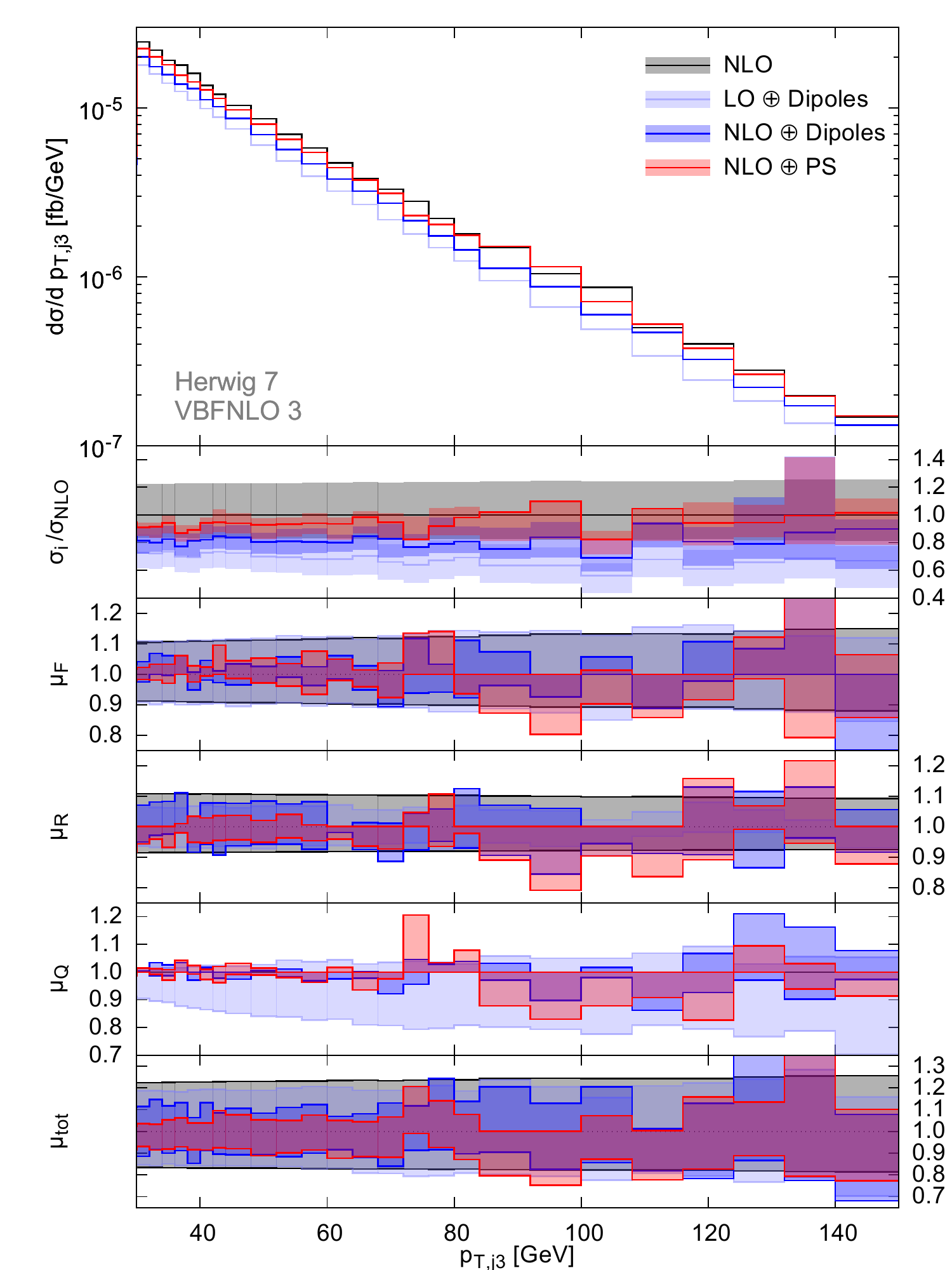}
\end{center}
\caption{\label{figures:Jet3Pt}The differential cross section for the
  transverse momentum of the third leading jet. See Fig.~\ref{figures:WWMass}
  and the text for more details.}
\end{figure}

\begin{figure}
\begin{center}
\includegraphics[scale=0.55]{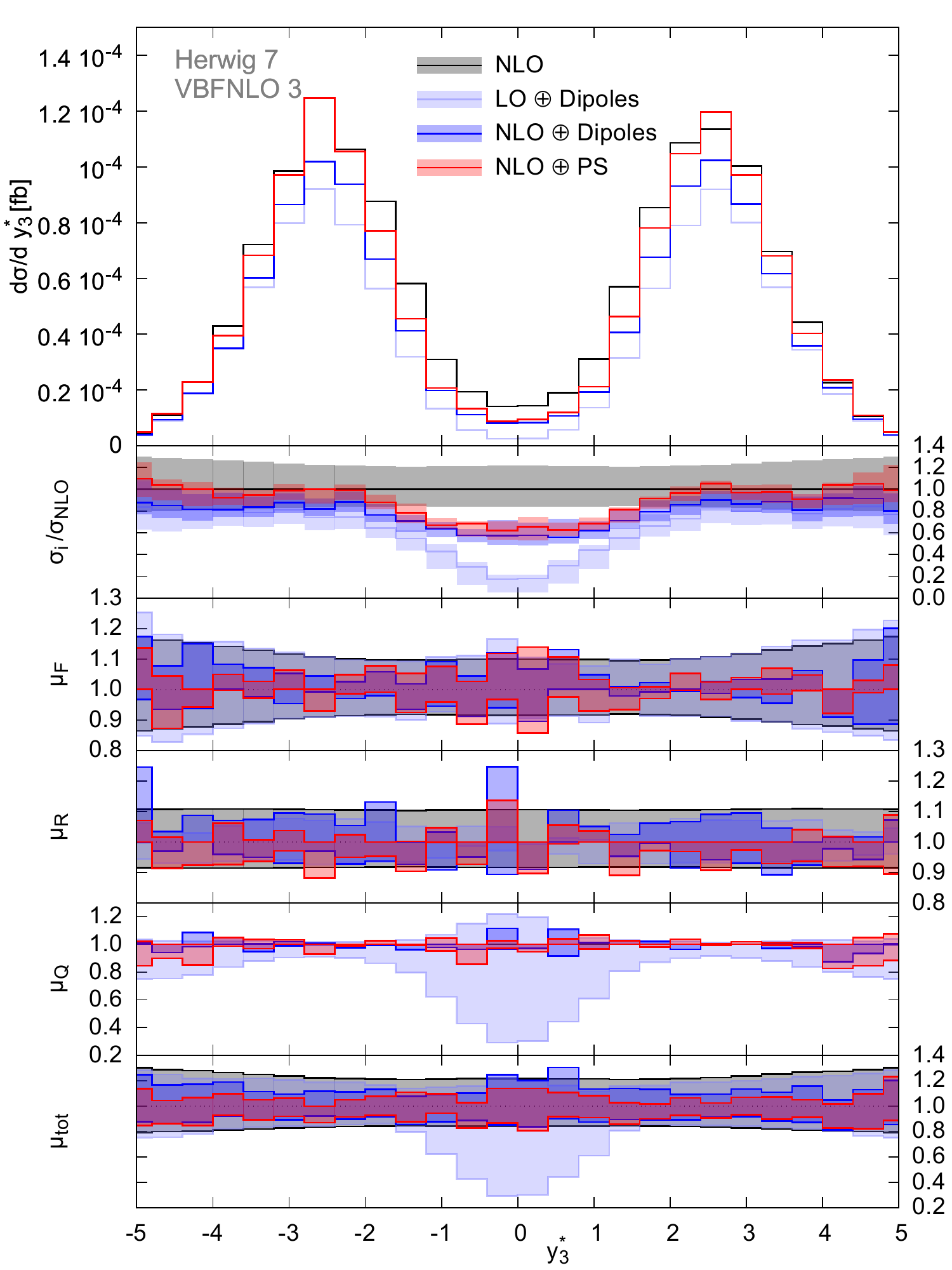}
\end{center}
\caption{\label{figures:YStar}The relative rapidity of the third leading jet
  with respect to the tagging jets. See Fig.~\ref{figures:WWMass} and the text
  for more details.}
\end{figure}

We finally turn to details of the third jet, as relevant to applying central
jet vetoes to suppress the impact of QCD-induced contributions. Since this
jet is present at leading order only in the matched simulation and solely
consists of parton shower radiation for the LO+PS setting, larger
uncertainties and impact of showering are expected. While small transverse
momenta of the third jet are, at NLO+PS, mostly stable with respect to shower
effects, Fig.~\ref{figures:Jet3Pt}, further details of the radiation pattern,
particularly the relative position of the third jet with respect to the
tagging jets\footnote{We use the 'un-normalised' definition, $y_3^* = y_3 -
  (y_1+y_2)/2$.}, Fig.~\ref{figures:YStar}, are significantly affected by both
the impact of NLO versus LO and additional shower emissions, as well.

\begin{figure}
\begin{center}
\includegraphics[scale=0.55]{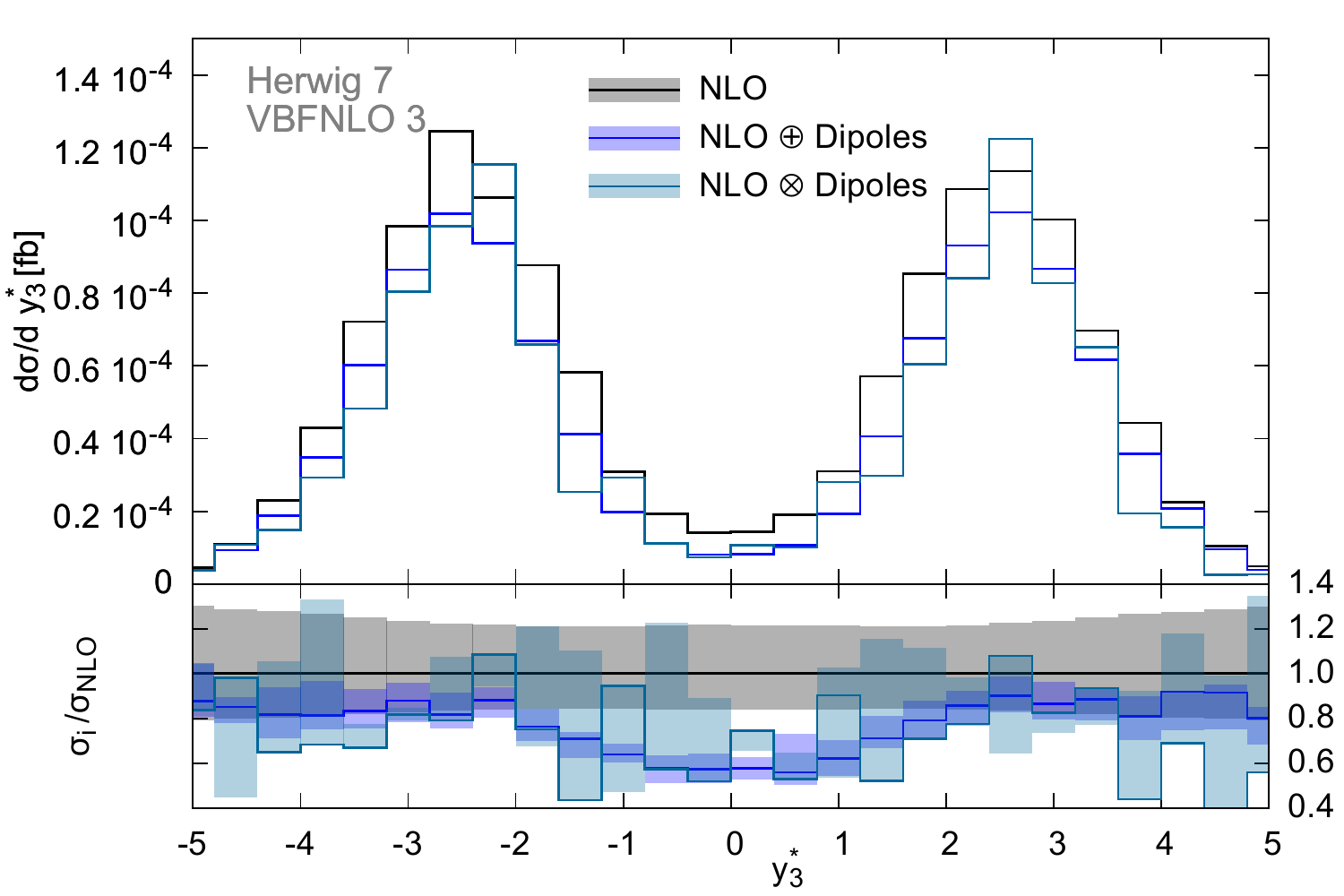}
\end{center}
\caption{\label{figures:PowhegCompare}The relative rapidity of the third jet
  with respect to the two tagging jets, comparing the fixed order parton level
  result (black) to the results obtained with the subtractive (blue) and
  multiplicative (dark cyan) matching algorithms. The ratio plot shows the
  ratio with respect to the fixed order result. The band denotes the
  change of the differential cross section when varying the
  factorization, renormalization and hard veto scale jointly around the
  central value, $\mu_F/\mu_0=\mu_R/\mu_0=\mu_Q/\mu_0 \in [\frac12;2]$.}
\end{figure}

At leading order we observe, for these observables, a large dependence on the
shower hard scale $\mu_Q$, which is reduced in the matched simulation though
still showing a deviation from the next-to-leading order shape for very
central jets in between the tagging jets. One would therefore be worried about the
choice of matching scheme, however, using a multiplicative (Powheg-type)
matching with a reasonable restriction on the exponentiated phase space by
applying the \texttt{resummation} profile scale, we find results compatible
with the subtractive matching, cf. Fig.~\ref{figures:PowhegCompare}.

\section{Conclusions and Outlook}

We have presented a study of NLO QCD predictions for electroweak $WW$ plus two
jet production including leptonic decays, off-shell effects and
non-resonant contributions. The fixed-order results have been
matched to subsequent parton showering using the two shower modules and the
\matchbox\ framework of \Hw~7, which has also been used to obtain the fixed
order results using amplitudes which have been made available via an extended
BLHA interface included in \vbfnlo~3.

Concentrating on perturbative physics at parton level, we find that matching
and parton shower uncertainties are well under control for this process. Given
that the third jet is described only at leading order, and higher jet
multiplicities are solely obtained from parton-shower radiation, we argue that
multi-jet merging in this case is desirable to further reduce the
uncertainties. Cut migration effects seem to impact the predictions at least
at the level of $10 \%$ and so require further investigation by {\it e.g.}
using vanishing generation cuts on jets and applying a reweighting procedure
to obtain sufficient statistics within the acceptance of the analysis.

As opposed to uncertainties at the level of the hard process and parton
showering, no consistent prescription has yet been obtained to assign
uncertainties to the overall event generator prediction including
hadronisation and multiple partonic interactions (MPI), which we leave for a
future study. The present work and tools used in it also constitute an
important contribution to a comprehensive programme of employing precision QCD
event generators for Higgs phenomenology in the VBF channel.

\section*{Acknowledgments}

We are grateful to the other members of the Herwig and VBFNLO collaborations
for encouragement and helpful discussions; in particular we would like to
thank Johannes Bellm, Stefan Gieseke and Peter Richardson for a careful review
of the manuscript. SP acknowledges support by a FP7 Marie Curie Intra European
Fellowship under Grant Agreement PIEF-GA-2013-628739, and the kind hospitality
of KIT on various occasions.

\bibliography{vbf-ww}

\end{document}